\documentclass{mn2e}

\usepackage[totalwidth=520pt, totalheight=680pt,letterpaper]{geometry}

\usepackage{times}
\usepackage[dvips]{graphicx}
\DeclareGraphicsExtensions{.ps}

\newcommand{\etal}{{et~al.}}

\newcommand{\bfx}{\bmath{x}}
\newcommand{\bfr}{\bmath{r}}
\newcommand{\bfR}{\bmath{R}}

\newcommand{\bfg}{\bmath{g}}

\newcommand{\bfU}{\bmath{U}}

\newcommand{\calO}{{\cal O}}

\newcommand{\bc}{\begin{center}}
\newcommand{\ec}{\end{center}}
\newcommand{\be}{\begin{equation}}
\newcommand{\ee}{\end{equation}}
\newcommand{\bea}{\begin{eqnarray}}
\newcommand{\eea}{\end{eqnarray}}
\newcommand{\befi}{\begin{figure}}
\newcommand{\enfi}{\end{figure}}
\newcommand{\befigw}{\begin{figure*}}
\newcommand{\enfigw}{\end{figure*}}

\newcommand{\zD}{z_{\scriptscriptstyle \mathrm{D}}}
\newcommand{\zG}{z_{\scriptscriptstyle \mathrm{G}}}

\newcommand{\tE}{t_{\mathrm{e}}}
\newcommand{\tO}{t_{\mathrm{o}}}
\newcommand{\tX}{t_{\mathrm{X}}}
\newcommand{\betaE}{\beta_{\mathrm{e}}}
\newcommand{\betaO}{\beta_{\mathrm{o}}}
\newcommand{\aO}{a_{\mathrm{o}}}

\newcommand{\vE}{v_{\mathrm{e}}}
\newcommand{\vO}{v_{\mathrm{o}}}

\newcommand{\HE}{H_{\mathrm{e}}}
\newcommand{\qE}{q_{\mathrm{e}}}
\newcommand{\DE}{D_{\mathrm{e}}}
\newcommand{\RE}{R_{\mathrm{e}}}
\newcommand{\TE}{T_{\mathrm{e}}}
\newcommand{\PsiE}{\Psi_{\mathrm{e}}}
\newcommand{\PsiO}{\Psi_{\mathrm{o}}}
\newcommand{\lambdaE}{\lambda_{\mathrm{e}}}
\newcommand{\lambdaO}{\lambda_{\mathrm{o}}}

\title[The cosmological redshift]{The kinematic component of the
  cosmological redshift}

\author[Chodorowski]{Micha{\l} J.\ Chodorowski\thanks{E-mail:
    michal@camk.edu.pl} \\ Copernicus Astronomical Center, Bartycka
  18, 00--716 Warsaw, Poland }

\begin{document}

\maketitle

\begin{abstract}
It is widely believed that the cosmological redshift is not a Doppler
shift. However, Bunn \& Hogg have recently pointed out that to settle
properly this problem, one has to transport parallely the velocity
four-vector of a distant galaxy to the observer's position. Performing
such a transport along the null geodesic of photons arriving from the
galaxy, they found that the cosmological redshift is purely
kinematic. Here we argue that one should rather transport the velocity
four-vector along the geodesic connecting the points of intersection
of the world-lines of the galaxy and the observer with the
hypersurface of constant {\em cosmic time}. We find that the resulting
relation between the transported velocity and the redshift of arriving
photons is {\em not\/} given by a relativistic Doppler formula.
Instead, for small redshifts it coincides with the well known
non-relativistic decomposition of the redshift into a Doppler
(kinematic) component and a gravitational one. We perform such a
decomposition for arbitrary large redshifts and derive a formula for
the kinematic component of the cosmological redshift, valid for any
FLRW cosmology. In particular, in a universe with $\Omega_\mathrm{m} =
0.24$ and $\Omega_\Lambda = 0.76$, a quasar at a redshift $6$, at the
time of emission of photons reaching us today had the recession
velocity $v = 0.997c$. This can be contrasted with $v = 0.96c$, had
the redshift been entirely kinematic. Thus, for recession velocities
of such high-redshift sources, the effect of deceleration of the early
Universe clearly prevails over the effect of its relatively recent
acceleration. Last but not least, we show that the so-called {\em
  proper\/} recession velocities of galaxies, commonly used in
cosmology, are in fact radial components of the galaxies'
four-velocity vectors. As such, they can indeed attain superluminal
values, but should not be regarded as real velocities.

\end{abstract}


\section{Introduction}
\label{sec:intro}
A standard interpretation of the cosmological redshift in the
framework of the Friedman-Lema{\^\i}tre-Robertson-Walker (FLRW) models
is that it is an effect of the expansion of the Universe. This
interpretation is obviously correct since $1 + z = a(\tO)/a(\tE)$,
where $z$ is the value of the redshift, $a(t)$ is the scale factor of
the Universe and $\tE$ and $\tO$ are respectively the times of
emission and observation of a sent photon. In semi-popular literature
(e.g. Kaufmann \& Freedman 1999; Franknoi, Morrison \& Wolff 2004;
Seeds 2007), but also in professional (e.g. Harrison 2000; Abramowicz
\etal\ 2007), one can often find statements that distant galaxies are
`really' at rest and the observed redshift is caused by the `expansion
of space'. According to other authors (e.g. Peacock 1999; Whiting
2004; Chodorowski 2007a,b; Bunn \& Hogg 2009, hereafter BH9), such
statements are misleading and cause misunderstandings about the
cosmological expansion. A presentation of the ongoing debate in the
literature on this issue is beyond the scope of the present work; a
(possibly non-exhaustive) list of papers includes Davis \& Lineweaver
(2001), Davis \& Lineweaver (2004), Barnes \etal\ (2006), Francis
\etal\ (2007), Lewis \etal\ (2007), Lewis \etal\ (2008), Gr{\o}n \&
Elgar{\o}y (2007), Peacock (2008), Abramowicz \etal\ (2009),
Chodorowski (2008), Cook \& Burns (2009).

On the other hand, there is broad agreement that the cosmological
redshift is not a pure Doppler shift; the gravitational field must
also generate a gravitational shift. Gravity can be neglected only
locally, in the local inertial frame (LIF) of an observer. It turns
out that for small redshifts, the cosmological redshift can be
decomposed into a Doppler shift and a Newtonian gravitational one
(Bondi 1947). The latter is a shift induced by the Newtonian
gravitational potential. Can the cosmological redshift be decomposed
into a Doppler shift and a gravitational shift (not necessarily
Newtonian) for an {\em arbitrary\/} value of the redshift? This is the
question which we want to deal with in this Paper. Formally, the
answer is no. There is no invariant definition of the recession
velocity of a distant galaxy in general relativity (GR). This velocity
is a relative velocity of the galaxy and the observer, and in curved
spacetime there is no unique way to compare vectors at widely
separated points. A natural way to define the recession velocity is to
{\em transport parallely\/} the velocity four-vector of the distant
galaxy to the observer, but the result will depend on the chosen
path. (This is just the definition of curvature.) In practice,
however, as a `preferred' path one can choose a {\em geodesic\/}
connecting the galaxy and the observer. Moreover, in FLRW models there
is a natural foliation of spacetime, into space-like hypersurfaces of
constant {\em cosmic time}. In our Paper, as the geodesic we will
adopt the geodesic lying on such a hypersurface, i.e.\ connecting the
points of intersection of the world-lines of the galaxy and the
observer with the hypersurface of constant cosmic time.

In a seminal study of the cosmological redshift, BH9, following
Synge~(1960) and Narlikar~(1994), adopted another geodesic for the
parallel transport: the null geodesic along which the photon is
travelling from the source to the observer. This approach results in
one `effective' velocity, while we think it is important to make a
distinction between the velocity {\em at the time of emission\/} and
the velocity {\em at the time of observation}. These two velocities
are obtained by transporting parallely the velocity four-vector of
the source respectively on the hypersurface of constant $\tE$ and
$\tO$. Not surprisingly, our result differs from that obtained by
BH9. However, unlike theirs, ours correctly reproduces the
small-redshift decomposition of the cosmological redshift into a
Doppler component and a gravitational component, mentioned
above. Specifically, our decomposition coincides with that of Bondi
(1947) for isotropic {\em and\/} homogeneous matter distribution.

This Paper is organized as follows. In Section~\ref{sec:transport} we
transport parallely the velocity four-vector of a distant galaxy to
the observer. From the transported vector we calculate the recession
velocity of the galaxy, which turns out to depend on the galaxy's
comoving distance and the assumed background cosmological model. In
Section~\ref{sec:specific} we find specific relations between the
cosmological redshift and its Dopplerian component for two
particularly simple cosmological models: the empty model and the
Einstein-de Sitter model. In Section~\ref{sec:PoE}, using only the
Principle of Equivalence, we derive the small-redshift decomposition
of the cosmological redshift and find it to be identical with that
obtained in Section~\ref{sec:specific} using generally relativistic
approach. In Section~\ref{sec:rec} we calculate the recession
velocities in the two models mentioned above, as well as for the
currently favoured, flat non-zero $\Lambda$ model with
$\Omega_{\mathrm{m}} = 0.24$. We find also that the velocities are
{\em subluminal\/} in {\em all\/} FLRW cosmological models and for
{\em all\/} values of the redshift. A comparison of the results
obtained in this Paper with some earlier works on the subject is given
in Section~\ref{sec:disc}. Summary is presented in
Section~\ref{sec:summ}.

\section{Parallel transport}
\label{sec:transport}
\textrm The equation for parallel transport of a vector $\bfU$ along a
curve with a parameter $\lambda$ and a tangent vector $\bfU = \{d
x^{\mu}\!/ d\lambda\}$ is

\begin{equation}
\frac{d U^\alpha}{d \lambda} + \Gamma^\alpha_{\,\mu \beta} \,\frac{d
  x^\mu}{d \lambda} U^\beta = 0 \,.
\label{eq:par-trans-christoff}
\end{equation}
Here, $\Gamma^\alpha_{\,\mu \beta}$ are the Christoffel symbols, related
to the underlying metric by the equation

\begin{equation}
\Gamma^\alpha_{\,\mu\beta} = {\textstyle \frac{1}{2}}
g^{\alpha\gamma} \!\left(g_{\gamma\beta,\mu} + g_{\gamma\mu,\beta} -
g_{\mu\beta,\gamma}\right) .
\label{eq:Gamma}
\end{equation}

The metric considered here is the
Friedman-Lema{\^\i}tre-Robertson-Walker (FLRW) metric. In the RW
coordinates, it is

\begin{equation}
ds^2 = c^2 dt^2 - a^2(t)[dx^2 + R_0^2 S^2(x/R_0) d\psi^2] \,.
\label{eq:RW}
\end{equation}
Here,
\be
d\psi^2 = d\theta^2 + \sin^2\!\theta\, d\phi^2 ,
\label{eq:dOmega}
\end{equation}
and $k R_0^{-2}$ is the present curvature of the universe. For a
closed, flat, and open universe, respectively, $k = +1, 0, -1$, and
the function $S(x)$ equals $\sin(x)$, $x$, and $\sinh(x)$. Time $t$ is
the cosmic time, that is the proper time of all Fundamental Observers
partaking in the homogeneous and isotropic expansion of the cosmic
substratum (fluid). The scale factor $a(t)$ relates fixed, or
comoving, coordinates, $\bfx$, to physical, or proper, coordinates,
$\bfr$: $\bfr = a \bfx$. The RW coordinates are $\{x^\mu\} = \{c t, x,
\theta, \phi\}$.

\begin{figure}
  \centering
  \includegraphics[width = \columnwidth]{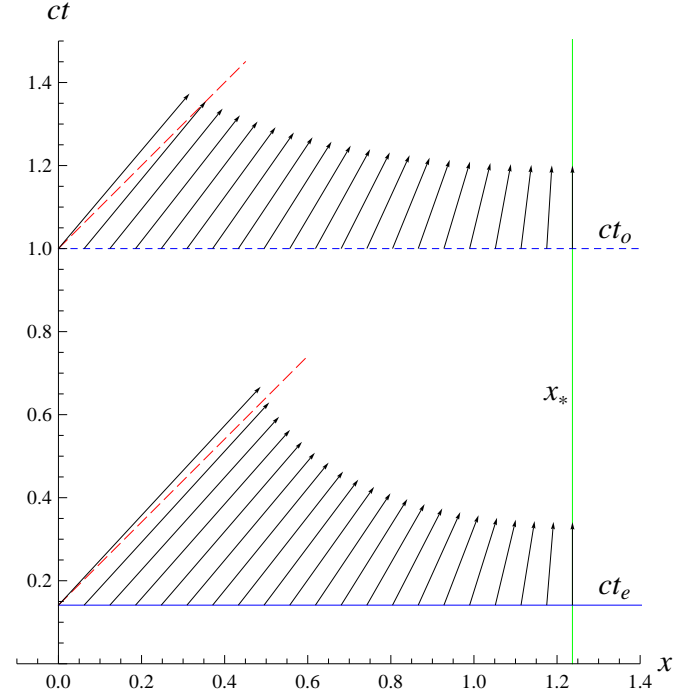}
  \caption{Parallel transport of the velocity four-vector of a distant
    galaxy, $\bfU$, from the galaxy to the observer along the geodesic
    on the hypersurface of constant cosmic time. The transport is
    plotted for two values of time: the observation time, $\tO$
    (dashed horizontal line), and the emission time, $\tE$ (solid
    horizontal line), both in units of $\tO$. The observer is in the
    origin of the coordinate system, $x = 0$, and the comoving
    coordinate of the emitting galaxy is $x_\ast$ (vertical line) in
    units of $c H_0^{-1}$. The transport is performed in comoving
    coordinates; in these coordinates all spatial components of the
    galaxy's velocity four-vector are initially zero. During the
    transport towards the observer, the radial component of the
    four-velocity gradually increases. For better visibility and
    reasons explained in Subsection~\ref{sub:specific}, in the plot
    this component is multiplied by $a(t)$. Slanted long-dashed lines
    show the direction of null vectors at the events $(\tO,0)$ and
    $(\tE,0)$ in {\em local inertial coordinates\/} of the observer. 
          }
  \label{fig:transport}
\end{figure}

The curve considered here is a radial geodesic on the hypersurface of
constant cosmic time, hence $d x^\mu = \delta^\mu_{~1} d x$, where
$\delta^\mu_{~\nu}$ is the Kronecker delta. This gives 

\begin{equation}
\frac{d U^\alpha}{d x} = - \Gamma^\alpha_{\,1 \beta} \,  U^\beta \,.
\label{eq:par-trans-radial}
\end{equation}
The vector $\bfU$ is here the four-velocity of a distant galaxy,
parallely-transported to the central observer (i.e.\ to the origin of
the coordinate system). By definition, $U^\alpha = dx^\alpha\!/ds$,
where $x^\alpha(t)$ denotes the galaxy world-line as a function of the
cosmic time. In the RW (comoving) coordinates every galaxy has fixed
$\theta$, $\phi$ and $x$, so $ds = c dt$ and $U^\alpha =
dx^\alpha\!/d(ct)$. For these reasons, the initial value for
$U^\alpha$ is $\delta^\alpha_{~0}$. Specifically, the initial
conditions for Equations~(\ref{eq:par-trans-radial}) are

\begin{equation}
U^\alpha(x_{\ast}) = \delta^\alpha_{~0} \,,
\label{eq:init}
\end{equation}
where $x_{\ast}$ is the comoving radial coordinate of the emitting
galaxy.

We calculate $\Gamma^\alpha_{\,1 \beta}$ using
equations~(\ref{eq:Gamma})--(\ref{eq:RW}) and obtain the following set of
two linear equations for $U^0$ and $U^1$:
\begin{eqnarray}
\frac{d U^0}{dx} &\!\!\! = \!\!\!& -\, \frac{a^2(t) H(t)}{c} \, U^1 ,
\label{eq:U^0} \\
\frac{d U^1}{dx} &\!\!\! = \!\!\!& -\, \frac{H(t)}{c} \, U^0 .
\label{eq:U^1}
\end{eqnarray}
We remind that since $t$ is fixed, so are $a$ and $H$. The
corresponding equations for $U^2$ and $U^3$ are simple to solve; with
conditions~(\ref{eq:init}), the result is $U^2(x) = U^3(x) =
0$. Differentiating Equation~(\ref{eq:U^0}) with respect to $r$ and
using~(\ref{eq:U^1}) to eliminate $d U^1\!/d x$ yields

\begin{equation}
\frac{d^2 U^0}{dx^2} = \frac{a^2 H^2}{c^2} \, U^0 .
 \label{eq:U^0-only}
\end{equation}
The solution is

\begin{equation}
U^0(x) = C_1 \mathrm{e}^{aHx/c} +
C_2 \mathrm{e}^{-aHx/c} ,
\label{eq:U^0-solution}
\end{equation}
where $C_1$ and $C_2$ are the integration constants. Imposing initial
conditions and noticing that $aH = \dot{a}$ we obtain
\begin{eqnarray}
U^0(x) &\!\!\! = \!\!\!&
{\textstyle \frac{1}{2}} \mathrm{e}^{\dot{a} (x - x_{\ast})/c} +
{\textstyle \frac{1}{2}} \mathrm{e}^{-\dot{a} (x - x_{\ast})/c} ,
\label{eq:U^0-sol} \\
U^1(x) &\!\!\! = \!\!\!&
- {\textstyle \frac{1}{2 a}} \mathrm{e}^{\dot{a} (x - x_{\ast})/c} +
{\textstyle \frac{1}{2 a}} \mathrm{e}^{-\dot{a} (x - x_{\ast})/c} ,
\label{eq:U^1-sol}
\end{eqnarray}
hence
\begin{eqnarray}
U^0(x=0) &\!\!\! = \!\!\!& \cosh{(\dot{a} x_{\ast}/c)} \,,
\label{eq:U^0-transported} \\
U^1(x=0) &\!\!\! = \!\!\!& a^{-1}\sinh{(\dot{a} x_{\ast}/c)} \,.
\label{eq:U^1-transported}
\end{eqnarray}
We see that $(g_{\alpha\beta} U^\alpha U^\beta) |_{x=0} = 1$, as it
should.

In order to identify the transported velocity four-vector of the
distant galaxy with its recession velocity, we now have to transform
the vector from the RW coordinates to the coordinates of the Local
Inertial Frame (LIF) of the central observer. A general coordinate
transformation $x^{\alpha'} = x^{\alpha'}\!(x^\alpha)$ transforms the
(unprimed) components of the metric, $g_{\alpha\beta}$, to

\begin{equation}
g_{\alpha'\beta'} = \frac{\partial x^\alpha}{\partial x^{\alpha'}}
\frac{\partial x^\beta}{\partial x^{\beta'}} \: g_{\alpha\beta} \,.
\label{eq:metric_transf}
\end{equation}
The following transformation of the RW coordinates: $t' = t, x' = a(t)x,
\theta' = \theta, \phi' = \phi$, yields $g_{\alpha'\beta'}$ which in
the limit $x' \to 0$ tend to $\eta_{\alpha'\beta'}$, i.e. to the
Minkowski metric.\footnote{To check this, instead of
using~eq.~(\ref{eq:metric_transf}) it is much simpler to note that $x =
x'/a(t')$ and to take its full differential.} Thus, the primed
coordinates are indeed the coordinates of the observer's LIF. A
transformation of vector components, $U^{\alpha'} = (\partial
x^{\alpha'} / \partial x^\alpha) U^\alpha$, yields here $U^{0'} = U^{0}$
and $U^{1'} = a(t) U^{1}$, hence
\begin{eqnarray}
U^{0'}(x'=0) &\!\!\! = \!\!\!& \cosh{(\dot{a} x_{\ast}/c)} \,,
\label{eq:U^0-transported_LIF} \\
U^{1'}(x'=0) &\!\!\! = \!\!\!& \sinh{(\dot{a} x_{\ast}/c)} \,.
\label{eq:U^1-transported-LIF}
\end{eqnarray}
Again, the components $U^{\alpha'}$ are properly normalized:
$(\eta_{\alpha'\beta'} U^{\alpha'} U^{\beta'})|_{x'=0} = 1$.

In the LIF of the observer, the radial component of the
parallely-transported four-velocity of the distant galaxy is
non-zero. We {\em interpret\/} this effect as a non-zero recession
velocity of the galaxy, $v$. Quantitatively, $U^{0'} = \gamma$ and
$U^{1'} = \beta\gamma$, where $\beta \equiv v/c$ and $\gamma \equiv (1
- \beta^2)^{-1/2}$. This yields immediately
\begin{eqnarray}
\gamma &\!\!\! = \!\!\!& \cosh{(\dot{a} x_{\ast}/c)}
\label{eq:gamma} \\
\beta\gamma &\!\!\! = \!\!\!& \sinh{(\dot{a} x_{\ast}/c)} \,,
\label{eq:beta-gamma}
\end{eqnarray}
hence\footnote{The quantity $\dot{a} x_{\ast}/c$ plays here
  essentially a role of the so called `velocity-parameter' of the
  Lorentz transformation.  See e.g.\ Rindler (1977) and Schutz
  (1985).}

\begin{equation}
\beta = \tanh{(\dot{a} x_{\ast}/c)} \,.
\label{eq:v_rec}
\end{equation}
Finally, since $\sinh{y} + \cosh{y} = \exp{y}$, we obtain
\begin{equation}
\exp\left(\frac{\dot{a} x_{\ast}}{c}\right) =
\left(\frac{1+\beta}{1-\beta} \right)^{1/2} \! = 1 + \zD \,,
\label{eq:redshift}
\end{equation}
where $\zD$ is the Dopplerian component of the (total) cosmological
redshift of the galaxy, $z$. In section~\ref{sec:specific} we will
explain how to use this equation to relate the cosmological redshift
of a distant galaxy to its Dopplerian component.
Equation~(\ref{eq:redshift}) is the central formula of the present
paper.

\section{Redshifts for specific models}
\label{sec:specific}
Let us remind that in Equation~(\ref{eq:redshift}), $x_{\ast}$ stands
for the comoving radial RW coordinate of a galaxy. For simplicity,
from now on we will denote this coordinate by $x$. In fact, $x$ is not
only a coordinate but also the comoving radial proper distance to the
galaxy and can be expressed in terms of the redshift of the
galaxy. For a given cosmological model, the relation between $x$ and
$z$ is determined by the Friedman-Lema{\^\i}tre (FL) equations. Using
this relation and Equation~(\ref{eq:redshift}) enables one to deduce
the relation between $z$ and $\zD$ in this model. In the following, we
will find the relation between $z$ and $\zD$ for two specific,
particularly simple cosmological models.

Formula~(\ref{eq:redshift}) is valid for any (fixed) value of cosmic
time. Two instants of time are of interest: the time of observation,
or `today', $\tO$, and the time of emission, $\tE$. The corresponding
values of the recession velocity are respectively $\vO = v(\tO)$ and
$\vE = v(\tE)$.

\subsection{Time of observation}
\label{subsec:t_o}
At the time of observation $\dot{a}|_{\tO} = H_0$ (we use the
normalization $\aO = 1$, so that the comoving radial distance equals
to the proper distance today), hence

\begin{equation}
\exp{\left[\frac{ H_0\, x(z)}{c}\right]} = 1 + \zD \,.
\label{eq:redshift-observ}
\end{equation}
Here, $\zD$ is a Doppler shift due to the {\em present\/} value of the
recession velocity, $v_\mathrm{o}$.

\subsubsection{Empty universe}
\label{subsub:empty-obs}
In the empty model ($\Omega_\mathrm{m} = \Omega_\Lambda = 0$), the
comoving radial distance is $x(z) = c H_0^{-1} \ln(1+z)$, hence
$\mathrm{e}^{\ln{(1+z)}} = 1+ \zD$, or

\begin{equation}
z = \zD \,.
\label{eq:redshift-empty-obs}
\end{equation}
In an empty universe the origin of the cosmological redshift is thus
entirely Dopplerian. This result is expected, since spacetime of an
empty universe is the Minkowski spacetime, where redshift is simply a
Doppler shift.

Some cosmologists still believe that the cosmological redshift is not
a Doppler shift {\em even\/} in the case of an empty universe. The
origin of this erroneous belief is a wrong interpretation of the FLRW
metric for an empty universe in the RW coordinates. In these
coordinates, space (3-D hypersurface of constant cosmic time) does
have non-zero curvature, but what matters here is the full 4-D
curvature, which is zero. Indeed, a simple coordinate transformation
transforms the metric to the standard Minkowskian form. What else
could spacetime of an empty universe be? There is no matter so there
is no gravity, and a complete relativistic description is provided by
special relativity. Indeed, Chodorowski (2007a) proved that in an
empty universe $z = \zD$, {\em using only specially relativistic
  concepts}. More specifically, he showed that $1+ \zD =
a(\tO)/a(\tE)$, where $a(t) = t/\tO$ is the scale factor of the empty
model. Another derivation of the redshift in the empty model will be
presented in Section~\ref{sec:PoE}.

There is some remaining subtlety here that deserves further
explanation. In the specially-relativistic formula for a Doppler
shift, $\zD = \sqrt{(1+v/c)/(1-v/c)}$, the recession velocity $v$ is
an inertial velocity of a distant galaxy, call it
$v_{\,\mathrm{distant}}$. The velocity that appears in
Equation~(\ref{eq:redshift}) and
resulting~(\ref{eq:redshift-empty-obs}) is a `local' velocity,
$v_{\,\mathrm{local}}$, transported parallely from the galaxy to the
central observer using the RW coordinates. Are these two velocities
equal?  Redshift is an observable, so the answer is `they must be',
but let us understand, why.

In Minkowski, i.e. flat, spacetime, {\em components\/} of a
transported four-vector in inertial coordinates are constant (see
Eq.~\ref{eq:par-trans-christoff}), so
$U^{\alpha''}_{~~~\mathrm{local}}$ (i.e., transported) is equal to
$U^{\alpha''}_{~~~\mathrm{distant}}$. Here, double primes denote the
global inertial coordinates of the central observer, in which
$v_{\,\mathrm{distant}}$ is measured. This equality is no longer true
for non-inertial coordinates (like RW), but parallel transport does
not depend on the coordinate system used, so the transported {\em
  vector\/} is the same. We have transformed the RW components of the
transported four-velocity, $U^{\alpha}_{~~~\mathrm{local}}$, to the
components in the coordinates of the LIF of the observer,
$U^{\alpha'}_{~~~\mathrm{local}}$, and obtained
$v_{\,\mathrm{local}}$. Since the global inertial frame of the central
observer is an extension of its LIF (possible only in an empty
universe), $U^{\alpha''}_{~~~\mathrm{local}}$ and
$U^{\alpha'}_{~~~\mathrm{local}}$ are the components of the same
vector in the same coordinate frame, so they must be equal. We thus
have $U^{\alpha''}_{~~~\mathrm{distant}} =
U^{\alpha''}_{~~~\mathrm{local}} = U^{\alpha'}_{~~~\mathrm{local}}$,
hence $v_{\,\mathrm{distant}} = v_{\,\mathrm{local}}$.

Summing up, our definition of the recession velocity of a distant
cosmological object correctly implies that in the case of an empty
universe the origin of the cosmological redshift is purely
Dopplerian.

\subsubsection{Einstein-de Sitter universe}
\label{subsub:EdS-obs}
In the Einstein-de Sitter (EdS) model ($\Omega_\mathrm{m} = 1$ and
$\Omega_\Lambda = 0$), the comoving radial distance is $x(z) = 2\, c
H_0^{-1}\, [1-(1+z)^{-1/2}]$, hence
\begin{equation}
2\left[1-(1+z)^{-1/2}\right] = \ln(1 + \zD) \,.
\label{eq:redshift-EdS}
\end{equation}
The particle horizon for the EdS model is $x(z = \infty) = 2\, c
H_0^{-1}$. We see that the Dopplerian component of the infinite
cosmological redshift of a source located at the horizon sphere is
$\zD = \mathrm{e}^2 - 1$. This finite value corresponds to the
recession velocity of the source {\em not\/} at the time of emission
of photons (i.e., $\tE = 0$), but {\em today\/}. In
sub-subsection~\ref{subsub:EdS-em} we will see that at $\tE = 0$, this
velocity was $c$. The difference between the two velocities is due to
deceleration of the EdS model.

\subsubsection{Small redshifts}
\label{subsub:redshift-small-obs}
A second-order expansion in redshift of the radial distance valid for
any cosmology (under the assumption that dark energy is in the form of
the cosmological constant) is $x(z) \simeq c H_0^{-1}\, [z -
  (1+q_0)z^2/2]$, where $q_0 \equiv \Omega_\mathrm{m}/2 -
\Omega_\Lambda$. (It is easy to check that second-order expansions of
the exact formulas for the empty and EdS models are consistent with
this general expansion.) From Equation~(\ref{eq:redshift}), after
expansion we obtain $\zD \simeq z - q_0\,z^2/2$, or

\begin{equation}
z = \zD + \frac{q_0}{2}\, \zD^2 + \calO\left(\zD^3\right) .
\label{eq:redshift-small-obs}
\end{equation}
In Section~\ref{sec:PoE} we will see that the above result is
identical to that obtained using the Principle of Equivalence (PoE).

\subsection{Time of emission}
\label{subsec:t_e}

At the time of emission, $\dot{a}|_{\tE} =
(\dot{a}/a)|_{\mathrm{e}} \,
a_{\mathrm{e}} = H_{\mathrm{e}} / (1
+ z)$, where $H_{\mathrm{e}}$ is the Hubble constant
at the time of emission. Using this equality in
Formula~(\ref{eq:redshift}) we obtain

\begin{equation}
\exp\left[\frac{ H(z)\, x(z)}{c\, (1+z)}\right] = 1 + \zD \,.
\label{eq:redshift-emission}
\end{equation}
Writing the above we have used the fact that there is a unique
correspondence between $\tE$ and $z$, so
$H_{\mathrm{e}}$ can be expressed in terms of the
latter. A general formula for the Hubble constant as a function of
redshift is

\begin{equation}
H^2(z) = H_0^2 \left[\Omega_\Lambda + \Omega_{\mathrm{m}} (1+z)^3 -
  (\Omega -1) (1+z)^2\right] .
\label{eq:H(z)}
\end{equation}
Here, $\Omega = \Omega_{\mathrm{m}} + \Omega_\Lambda$. In
Equation~(\ref{eq:redshift-emission}), $\zD$ is a Doppler shift due to
the recession velocity at the time of {\em emission\/},
$v_{\mathrm{e}}$.

\subsubsection{Empty universe}
\label{subsub:empty-em}
From Equation~(\ref{eq:H(z)}) we obtain $H(z) = H_0 (1+z)$, hence
$H(z) x(z)/[c (1+z)] = \ln{(1+z)}$. This yields again

\begin{equation}
z = \zD \,.
\label{eq:redshift-empty-em}
\end{equation}
Equations~(\ref{eq:redshift-empty-em})
and~(\ref{eq:redshift-empty-obs}) involve two different Doppler
shifts: respectively at the time of emission, $z_{\scriptscriptstyle
  \mathrm{D,e}}$, and at the time of observation, $z_{\scriptscriptstyle
  \mathrm{D,o}}$. These equations are consistent because in the empty
model the expansion of the universe is kinematic and recession
velocities are constant in time. This is not true for any other
cosmological model.

\subsubsection{Einstein-de Sitter universe}
\label{subsub:EdS-em}
In the EdS model $H(z) = H_0(1+z)^{3/2}$, hence

\begin{equation}
\zD = \exp{\left\{2\!\left[(1+z)^{1/2} - 1\right]\right\}} - 1 \,.
\label{eq:redshift-EdS-em}
\end{equation}
Infinite value of $z$ corresponds to infinite value of $\zD$, so at
the time of the Big-Bang (corresponding to $z = \infty$), the horizon
was receding from the observer with the velocity of light. Let us
decompose the cosmological redshift into a Doppler shift and the
`rest', $z = \zD + \mathrm{`rest}$'. From
Equation~(\ref{eq:redshift-EdS-em}) we see that for high redshifts,
the Dopplerian component of the cosmological redshift is (much) bigger
than the total value of the latter.  Therefore, in the EdS model the
`rest' must in fact be negative, i.e., it is a blueshift. We will show
later that for small redshifts, the `rest' can be identified with a
gravitational shift. It will be demonstrated that a gravitational
shift is indeed a blueshift for {\em all\/} matter-only models. A
homogeneous and isotropic distribution of ordinary matter creates a
gravitational potential well with the observer at the center and the
work performed by photons climbing {\em down\/} this well is negative.

\subsubsection{Small redshifts}
\label{subsub:redshift-small-em}

From Equation~(\ref{eq:H(z)}) we have $H(z) = H_0 [1+(1+q_0)z
+ \calO(z^2)]$, hence $H(z)\, x(z)/[c (1+z)] \simeq z + (q_0-1)
z^2/2$. Inserted in~Equation~(\ref{eq:redshift-emission}), this yields

\begin{equation}
z = \zD - \frac{q_0}{2}\, \zD^2 + \calO\left(\zD^3\right) .
\label{eq:redshift-small-em}
\end{equation}
Again, in Section~\ref{sec:PoE} we will see that the above result is
identical to that obtained using the PoE.

\section{Principle of Equivalence}
\label{sec:PoE}

In Section~\ref{sec:specific} we studied the cosmological redshift
using generally-relativistic concepts. In this Section we will
restrict our analysis to small values of the cosmological redshift;
this will allow us to use the Principle of Equivalence (PoE). It is
well known that the PoE enables one to study local effects of light
propagation in a weak gravitational field. Specifically, one can
replace a static reference frame in this field by a freely falling
frame in empty space. From the PoE it follows that the latter is
locally inertial, with the laws of light propagation being given by
special relativity. The classical examples of phenomena studied in
this way are the gravitational redshift and the bending of light
rays. In some instances it is possible to calculate correctly
non-local effects by summing up infinitesimal contributions from
adjacent local inertial frames, freely falling along light ray
trajectory. For example, the (small) gravitational redshift of a
photon sent from the surface of a non-relativistic star (like, e.g.,
the Sun) to infinity can be calculated in this way. In some other
instances it is not possible; a classical example is the bending of
light rays. The calculation based on the PoE gives only half of the
value of bending predicted by GR in the weak-field approximation (but
see Will 2006 and references therein).

It is well known that the cosmological redshift can be interpreted as
`an accumulation of the infinitesimal Doppler shifts caused by photons
passing between fundamental observers separated by a small distance'
(Peacock 1999; see also Peebles 1993, Padmanabhan 1993). The
derivation is very simple but let us recall it here. Consider two
neighbouring comoving observers, located on the trajectory of a light
ray, separated by an infinitesimal proper distance $\delta l = a(t)
\delta x$, where $a(t)$ is the scale factor and $\delta x$ is their
comoving separation. The velocity of the second observer relative to
the first is $\delta v = (d/dt) \delta l = \dot{a} \delta x$. The
equation for the radial light propagation in the RW metric is $c\:\!
\delta t = a \delta x$, where $\delta t$ is the light travel time
between the two observers, so $\delta v = c\:\!  \delta a / a$. This
small recession velocity of the second observer causes a
non-relativistic Doppler shift of the passing ray, $\delta \lambda /
\lambda = \delta v / c = \delta a / a$. (Reference frames of the FOs
are locally inertial.)  Integrating the relation between $\lambda$ and
$a$ we obtain $1 + z \equiv \lambda_o / \lambda_e = a_o/a_e$.

The above derivation resembles a derivation using the PoE, which will
be presented below. There is, however, an important difference. In the
above we have applied the generally relativistic RW metric and the law
of light propagation in it. The RW coordinates are local inertial
coordinates of (freely falling) Fundamental Observers and the cosmic
time $t$ is their local proper time. This has enabled us to integrate
the effect (along the ray's trajectory) trivially and exactly. Not
surprisingly, then, the result is correct for an arbitrary value of
the redshift.  In an application of the PoE, the gravitational field
of the system under consideration is described in terms of
non-relativistic Newtonian dynamics, so the field must be weak. In the
FLRW cosmological models the gravitational field is weak only at small
distances from an observer, more specifically much smaller than the
Hubble radius. Therefore, the PoE can give correct predictions only
for small values of the cosmological redshift. Still, we think that
calculations using the PoE are worth performing. The reasons are as
follows.

First, the aim of our Paper is to decompose the cosmological redshift
into a {\em global\/} Doppler shift (i.e., due to the relative
velocity of the emitter and the observer) and the `rest'. In the above
calculation of the redshift we summed up contributions from
infinitesimal relative velocities of neighbouring FOs, but this did
not help us to express the answer in terms of the relative velocity of
the emitter and the observer. As stated already many times, for large
cosmological distances the recession velocity a priori is not well
defined. In this Paper we proposed its generally relativistic
definition. This definition is however quite formal, while for small
redshifts, velocity is a well defined pre-relativistic concept.
Therefore, the identification of a global Doppler component of the
small cosmological redshift will be simple and unambiguous. Second, in
previous sections, using GR, we have shown that the cosmological
redshift, of any value, can be indeed decomposed into a Doppler shift
and the `rest'. In this section, using the PoE, we will similarly
decompose the redshift into a global Doppler shift and the rest and
will show that the `rest' is the gravitational redshift in the
Newtonian sense. In other words, we will prove that for small
redshifts the rest is a classical Newtonian gravitational shift.

Among all the FLRW models there is a toy model which can be described
using special relativity only. This is the empty, or Milne model
(Milne 1933). At first sight, the relation of this model to the PoE is
problematic, since in an empty universe there is no gravity. However,
the relevance of the model will be seen later. Spacetime of an empty
universe is the Minkowski spacetime. Let us denote the {\em global}
inertial coordinates of an observer by $T$, $X$, $Y$ and $Z$. In the
standard spherical coordinates, $R$, $\Theta$, $\Phi$, the metric is

\begin{equation}
ds^2 = dT^2 - dR^2 - R^2 (d\Theta^2 + \sin^2\Theta\, d\Phi^2) \,.
\label{eq:Mink}
\end{equation}
Like above, we will calculate the cosmological redshift summing up the
infinitesimal Doppler shifts caused by photons passing between
neighbouring Fundamental Observers, located along the photons' path.
We have seen that such a summation can be easily done in the RW
coordinates, for any FLRW model. However, it turns out that in the
empty model, it is possible to perform this summation using also the
Minkowskian coordinates $T$ and $R$.

In the Milne model the recession velocities of Fundamental Observers
are constant; for example, in the rest-frame $O$ of the central
observer $v(T,R) = R/T = H(T) R = \mathrm{const}$. Let us consider a
Fundamental Observer located at a distance $R$ from the central
observer at time $T$. In his rest-frame $O'$, he observes a small
Doppler shift of a photon sent (in fact, transmitted) by a nearby
observer located at $R + \delta R$. Due to the Hubble expansion, the
latter observer has a small recession velocity $\delta v' = \delta
R'/T' = H(T')\delta R'$ relative to the former.  Like in the RW
coordinates (locally coinciding with the coordinates $T'$ and $R'$),
the relative wavelength shift is $\delta \lambda/\lambda = \delta
v'/c$. However, in the empty model we can use the special-relativistic
law of composition of velocities to relate the velocity $\delta v'$ to
the velocity $\delta v$, measured in the observer's frame $O$. This
law can be applied, because the relative velocity of the frames $O$
and $O'$, $v(T,R)$, is constant. We have

\begin{equation}
\delta v' = \frac{v(T,R+\delta R) - v(T,R)}{1 - v(T,R+\delta R)\,
  v(T,R)/c^2} = \frac{\delta v(T,R)}{1 - \left[v(T,R)/c\right]^2} \,,
\label{eq:vel-rel}
\end{equation}
hence we obtain

\begin{equation}
\frac{\delta \lambda}{\lambda} = - \frac{\delta v(T,R) / c}{1 -
  \left[v(T,R)/c\right]^2} \,.
\label{eq:lambd-rel}
\end{equation}
Let us denote the distance of the source at the time of emission by
$\DE$. Our present task is to integrate this equation along the
photon's trajectory, $c(T - \TE) = \DE - R$. (The photon travels {\em
  towards} the central observer.) However, it can be easily checked
that since the velocity of any FO is constant, $\delta v(T,R) =
H(T)\delta R$ equals to $\delta v_e = H(\TE)\delta \RE$. Therefore,
the integration simplifies to an integration over $\RE$ at fixed $T =
\TE$, 

\begin{equation}
\int_{\lambdaE}^{\lambdaO} \frac{d \lambda}{\lambda} =
- \int_{\DE}^{0} \frac{(\HE/c)\, d \RE}{1 - (\HE \RE/c)^2} \,.
\label{eq:int-Milne1}
\end{equation}
(The minus sign accounts for the fact that for a photon travelling
from the source to the observer $R$ decreases, so $\delta R$ is
negative.) Defining $\beta \equiv \HE \RE / c$ and $\betaE \equiv \HE
\DE / c$, we obtain

\begin{equation}
\int_{\lambdaE}^{\lambdaO} \frac{d \lambda}{\lambda} =
\int_0^{\betaE} \frac{d\beta}{1 - \beta^2} \,.
\label{eq:int-Milne}
\end{equation}
The left-hand-side integral is equal to $\ln(1+z)$. The integration of
the right-hand-side is elementary and yields
$\ln[(1+\betaE)/(1-\betaE)] /2$. Hence,

\begin{equation}
1 + z = \left(\frac{1 + \betaE}{1 - \betaE}\right)^{1/2} = 1 + \zD .
\label{eq:redshift-Milne-SR}
\end{equation}
In sum, we have accumulated infinitesimal Doppler shifts using the
{\em inertial\/} coordinates of the central observer and expressed the
total redshift in terms of the recession velocity of the source. The
relation between the cosmological redshift and the recession velocity
turned out to be given by the relativistic Doppler formula, in
agreement with generally-relativistic
Equation~(\ref{eq:redshift-empty-em}). Thus, as already stated 
earlier, in an empty universe the cosmological redshift is entirely a
global Doppler shift. A similar derivation of the redshift in the
empty model using Minkowskian coordinates with an {\em explicit}
integration along the photon's trajectory was presented in Appendix~A
and Section~2 of Chodorowski (2007a).

Now we proceed to derive, using the PoE, the relation between the
recession velocity of a nearby source and its redshift, for {\em
  non-empty} universes. Why was the exercise with the Milne model
useful for this purpose? High redshifts correspond to relativistic
velocities and in the empty model their effects manifest as the $1 -
\beta^2$ factor in the integrand on the RHS of
Equation~(\ref{eq:int-Milne}). Expansion of this integrand for small
velocities and subsequent integration yields $\betaE +
\calO(\betaE^3)$.  Therefore, in our non-relativistic calculations we
will be allowed to retain terms up to second order in $\betaE$. (We
will see later that the leading-order contribution to the
velocity--redshift relation coming from Newtonian gravity is already
of the second order in $\betaE$.) Like previously, we denote Newtonian
coordinates of the central observer by $T$ and $R$. They are now
inertial only locally, because there is a gravitational field. We
`transform this field away' using local inertial frames of
FOs. Contrary to the empty model, FOs do not move now with constant
speeds. According to the Newtonian interpretation of the FL equations
(Milne \& Mc Crea 1934), the change of speed of any FO located on a
sphere centered on another observer is induced by the gravitational
force exerted on the FO by the mass encompassed within this
sphere. (This is Newton's theorem for an isotropic and homogeneous
universe. The speed of the FO is, of course, measured relative to the
central observer). The effects of the cosmological constant $\Lambda$
can be incorporated easily because $\Lambda$ results in a force
proportional to distance, similarly to the classical gravitational
force for a homogeneous mass distribution.  The FO located at time $T$
at a distance $R$ from the central observer has velocity related to
his velocity at the time of emission of a photon, $\vE$, in the
following way:

\begin{equation}
v(T,R) = \vE + g(\TE,\RE) \Delta T + \calO(\Delta T^2) \,.
\label{eq:small_velocity}
\end{equation}
Here, $\TE$ is the time of emission, $\RE = R(\TE)$, $\Delta T = T -
\TE$ and $g(\TE,\RE)$ is the initial gravitational acceleration of
the FO. Writing the above formula we have approximated changing
acceleration of the FO by its initial value. (The result of this
approximation for $v(T,R)$ is already of second order in $\Delta
T$.) It is straightforward to calculate $g(T,R)$ from Newton's
theorem and to express it in terms of the cosmological parameters.
However, there is a faster way of doing this: by differentiating the
Hubble law in Newtonian coordinates, $v = H(T) R$, with respect to
time. The result is

\begin{equation}
g(T,R) = - q(T) H^2(T) R \,, \label{eq:g(T,R)}
\end{equation}
where $q$ is the deceleration parameter.

From Equation~(\ref{eq:g(T,R)}), $g(\TE,\RE) = - \qE \HE^2 \RE \simeq
- q_0 \HE^2 \RE$. Inserting these expressions for $\vE$ and $g$ in
Equation~(\ref{eq:small_velocity}), we obtain up to terms linear in
$\Delta T$

\begin{equation}
v(T,R) = \HE \RE (1 - q_0 \HE \Delta T)
\label{eq:small_velocity_2}
\end{equation}
($R \simeq \RE + \HE \RE \Delta T$). The velocity of a FO is described
here in terms of $T$ and his initial position, $\RE$. A small relative
velocity of two neighbouring FOs at time $T$, $\delta v_{|T}$, can be
expressed as $(\partial v / \partial \RE)_{|T} \delta \RE = \HE (1 -
q_0 \HE \Delta T) \delta \RE$. This relative velocity causes a
non-relativistic Doppler shift of the frequency of a passing
photon. To obtain the total redshift we need to integrate these
infinitesimal Doppler shifts along the photon's path (i.e., null
geodesic). Unlike in the empty model, here $T$ and $R$ are only
locally inertial coordinates. However, their extent is quite limited,
$\DE \ll c \HE^{-1}$, so $\HE \Delta T$ equals to $\HE (\DE - R)/c
\simeq \HE (\DE - \RE)/c$ plus terms already of higher order in $(\DE
- \RE)/c$. Therefore, $\int_{\lambda_e}^{\lambda_o} d\lambda / \lambda
= \ln(1 + z)$ equals to

\begin{equation}
\frac{1}{c} \int_0^{\DE} \HE \left[ 1 - q_0 \frac{\HE (\DE - \RE)}{c}
\right] d \RE = \betaE - \frac{1}{2} q_0 \betaE^2 \,,
\label{eq:small_redshift_PoE}
\end{equation}
where, as before, $\betaE = \HE \DE /c$. Hence we obtain finally

\begin{equation}
z = \exp\!\left[\betaE - \frac{1}{2} q_0 \betaE^2\right] - 1 =
\betaE + \frac{1}{2} \betaE^2 - \frac{q_0}{2} \betaE^2 +
\calO(\betaE^3) \,.
\label{eq:small_redshift__final}
\end{equation}
Expansion of the Doppler formula~(\ref{eq:redshift-Milne-SR}) to
second order in $\betaE$ gives $\zD = \betaE + \betaE^2/2$. Therefore,
the first two terms on the RHS of
Equation~(\ref{eq:small_redshift__final}) can be identified with a
(global) Doppler shift due to the recession velocity of the source at
the time of emission. Since the third term is of second order in
$\betaE$, we can write $q_0 \betaE^2/2 = q_0 \zD^2/2 +
\calO(\zD^3)$. In such a way we find out that
Equation~(\ref{eq:small_redshift__final}) is identical to
Equation~(\ref{eq:redshift-small-em}), the latter derived using a
generally-relativistic approach. In
Equation~(\ref{eq:redshift-small-obs}), the kinematic component of the
redshift is related by the Doppler formula to the recession velocity
of the source at the time of {\em observation}. We recover this
equation by noting (with some small labour) that
Equation~(\ref{eq:small_velocity_2}) yields $\betaE = \betaO + q_0
\betaO^2$, and using this equality in
Equation~(\ref{eq:small_redshift__final}).

The term $- q_0 \betaE^2/2$ is a gravitational shift: the work, per
unit-mass rest energy, performed by a photon travelling from the
source to the observer. From Equation~(\ref{eq:g(T,R)}) we have $\bfg
\simeq - q_0 \HE^2 \bfR$, so the gravitational potential $\Psi(R)
\simeq q_0 \HE^2 R^2 /2 + \mathrm{const}$. We thus indeed obtain
$(\PsiO - \PsiE) /c^2 = - q_0 \betaE^2/2$. Whether the observer is
located in the centre of a potential well or on the top of a potential
hill depends on the value of $q_0$. For all matter-only cosmological
models $q_0$ is positive, so the observer is in the centre of a
potential well. Then a gravitational shift is negative, so it is in
fact a blueshift. This reflects the fact that the photon climbs {\em
  down\/} a potential well so the performed work is negative. For an
accelerating universe (like, as we currently believe, ours) $q_0 < 0$
and a gravitational shift is a redshift.

Peacock (1999) derived Equation~(\ref{eq:small_redshift__final}) in a
much simpler way. He stressed that in the definition of the redshift,
the wavelength at the time of emission is the wavelength measured in
the source's rest-frame $O'$, ${\lambdaE}'$. In the observer's
rest-frame $O$ this wavelength is Doppler-shifted,
$\lambdaE/{\lambdaE}' = 1 + \zD$. Next, on the path to the observer
the photon undergoes a gravitational shift, $\lambdaO/\lambdaE = 1 +
\zG$. Hence, $1 + z \equiv \lambdaO/{\lambdaE}' = (1 + \zD) (1 +
\zG)$. This equality gives Formula~(\ref{eq:small_redshift__final})
because, as we have verified above, $\zG = - q_0 \betaE^2/2$, and
because $\zD \zG$ is already of third order in $\betaE$. However,
freely-falling observers employed in his calculation of $z$ were {\em
  not\/}, strictly speaking, the freely-falling observers of the
studied problem. In the RW models freely-falling observers are (by
definition) FOs, partaking in the Hubble expansion. Peacock employed
the standard `Einstein-lift' observers, starting to fall only when the
photon passes the top of their cabin. Accounting for non-zero initial
velocities of freely-falling observers complicated our calculations,
but enabled us to apply the PoE `from first principles'. We performed
our calculations consistently in the same way as we did using global
inertial coordinates in the empty model, and the RW coordinates in all
FLRW models.

Peacock's aim was to derive a formula for the angular diameter
distance up to the terms quadratic in redshift. In order to obtain the
correct result, he had to expand the relativistic Doppler formula up
to the term quadratic in $\betaE$. Strictly speaking, this term is a
relativistic correction and in principle one is not allowed to account
for it in the calculations using the PoE, where velocities are assumed
to be non-relativistic. However, the justification for including this
term can be found analyzing our calculations in the Milne model. We
obtained $\ln(1 + \zD) = \betaE + \calO(\betaE^3)$, or $1 + \zD =
\exp[\betaE + \calO(\betaE^3)]$, where $\calO(\betaE^3)$ was the
lowest-order relativistic correction. Therefore, solving for $\zD$ we
were allowed to expand the exponential function up to a term quadratic
in $\betaE$, still remaining in the non-relativistic regime. Since the
exponential function has the same second-order expansion as the
Doppler formula, we correctly obtained the second-order Doppler
relation, also in our non-relativistic calculations for non-empty
models (see Eq.~\ref{eq:small_redshift__final}).

We stress that we did not {\em assume\/} that $z = \zD + \zG$. In our
calculations both terms appeared naturally as a result of an
application of the PoE.  The first person to note that the cosmological
redshift, if small, can be decomposed into a Doppler shift and a
gravitational shift was Bondi (1947). He performed such a
decomposition in a spherically symmetric generalization of the RW
models, the Lema{\^\i}tre-Tolman-Bondi model. Our
Formula~(\ref{eq:small_redshift__final}) is a special case of his
formula~(52), in a sense that $\zG = - q_0 \betaE^2/2$ only for a
homogeneous distribution of cosmic matter and small distances.

In the approach presented in this section, the cosmological redshift
of {\em any\/} value is generated by relative motions of FOs. For
small values of the redshift the application of the PoE allowed us to
decompose the redshift into a Doppler component and a gravitational
component.\footnote{Another derivation of this decomposition was
  presented in Gr{\o}n \& Elgar{\o}y (2007).} We have found that a
Doppler component of the redshift is obviously generated by a
kinematic component of the motions, while a gravitational component of
the redshift is generated by their varying-in-time component. Of
course, this interpretation of a gravitational component is fully
consistent with the standard interpretation of a gravitational shift,
because this is a gravitational field that causes FOs to accelerate.

\section{Recession velocities}
\label{sec:rec}

Calculations in Section~(\ref{sec:transport}), based on our definition of
recession velocity, led us to Equation~(\ref{eq:v_rec}). This equation
is a general formula for the recession speed of a galaxy, valid for
an arbitrary comoving position of the galaxy and an arbitrary instant of
time. The hyperbolic tangent on the RHS of this equation ensures that
the recession velocity of a source with an arbitrarily high redshift
is subluminal ($v < c$). However, it is widely believed that recession
velocities of sufficiently distant galaxies can be, and indeed are,
superluminal (e.g. Davis \& Lineweaver 2001; Davis \& Lineweaver 2004;
Gr{\o}n \& Elgar{\o}y 2007; Lewis \etal\ 2007). In
Subsection~(\ref{sub:superluminal}) we will find a relation between
our recession velocity and a commonly used estimator of the velocity:
the {\em proper} recession velocity. Basing on this relation we will
show that the conception of the superluminal expansion of the Universe
is wrong. In Subsection~(\ref{sub:specific}) we will intercompare our
recession velocities of galaxies for some specific cosmological
models.

\subsection{No superluminal velocities}
\label{sub:superluminal}

Let us remind that in the comoving coordinates $\{x^\mu\} =
\{ct,x,\theta,\phi\}$, the FLRW metric takes on the form given by
Equation~(\ref{eq:RW}). For a particle (or a galaxy) participating in
the homogeneous and isotropic expansion of the cosmic fluid, $\theta =
\mathrm{const}$, $\phi = \mathrm{const}$ and $x = \mathrm{const}$.
Hence, $ds = c dt$ and

\begin{equation}
U^{\alpha} = \frac{dx^{\alpha}}{c dt} = \left(1, 0, 0, 0 \right) .
\label{eq:U^alpha}
\end{equation}
Let us change coordinates to $\{x^{\mu'}\} = \{ct,r,\theta,\phi\}$,
where $r = a(t)x$. Spacetime interval $ds$ is an invariant, so also in
the primed coordinates $ds = c dt$. The four-velocity of the galaxy in
these coordinates is

\begin{equation}
U^{\alpha'} = \frac{dx^{\alpha'}}{c dt} = \left(1, \frac{H r}{c}, 0, 0
\right) ,
\label{eq:U^alpha'}
\end{equation}
where $H(t) = \dot{a}/a$ is the Hubble constant. 

It is a standard practice in the literature to use as an estimator of
the recession velocity of a galaxy its so-called {\em proper\/}
recession velocity,

\begin{equation}
v_\mathrm{prop} \equiv \frac{dr}{dt} = H r \,,
\label{eq:v_prop}
\end{equation}
that is the derivative of the proper distance between the galaxy and
the observer with respect to the proper cosmic time. Comparing
equation~(\ref{eq:U^alpha'}) with~(\ref{eq:v_prop}) we see that

\begin{equation}
v_\mathrm{prop} = c\, U^{1'} \,,
\label{eq:v_prop-our}
\end{equation}
or that $v_\mathrm{prop}/c$ is nothing more than the radial component
of the galaxy's four-velocity in the {\em proper\/} coordinates
$\{ct,r,\theta,\phi\}$. Therefore, the proper velocity can indeed
attain superluminal values,\footnote{For instance, in
  Fig.~\ref{fig:transport} the present value of the proper velocity of
  the source is $1.24 c$.} but should not be regarded as a recession
velocity. As described earlier, to calculate the latter, one should

\begin{enumerate}
  \item Transport parallely the galaxy's velocity four-vector to the
  observer's position;

  \item Transform the components of the transported vector to
    the components in the locally inertial coordinates of the
    observer;

  \item Read off the value of relative velocity using the fact that in
    these coordinates, $U^{1''} = \beta \gamma(v)$.
\end{enumerate}
We stress that $U^\alpha$ and $U^{\alpha'}$ given in
Equations~(\ref{eq:U^alpha}) and~(\ref{eq:U^alpha'}) are just
different components {\em of the same four-vector} $\bfU(t,r)$ in
respectively comoving and proper coordinates. It has been noted
earlier that in the vicinity of the observer, the proper coordinates
reduce to her/his local inertial coordinates. Therefore, in this case
the operation B.\ should not be necessary. Indeed, a straightforward
(though somewhat more complicated) calculation of the parallel
transport gives directly $U^{1'}(t,r=0) = \sinh{(H r/c)}$, in
agreement with Equation~(\ref{eq:U^1-transported-LIF}).

Note that for sufficiently distant objects, the transported (in proper
coordinates) radial component of the four-velocity is even more
superluminal than the initial one [$\sinh{(H r/c)}$ compared to
  $Hr/c$].\footnote{In the model used to construct
  Fig.~\ref{fig:transport}, the final value of the radial component
  transported on the hypersurface $t = \tO$ is $1.58$ compared to its
  initial value $1.24$.} However, using the equation given in point
C.\ of the above list, we get

\begin{equation}
\beta_\mathrm{our} = \tanh{(v_\mathrm{prop}/c)} \,,
\label{eq:v_rec-prop}
\end{equation}
where $v_\mathrm{prop} = H r$. Equation~(\ref{eq:v_rec-prop}), which
coincides with equation~(\ref{eq:v_rec}), relates the proper recession
velocity of a galaxy to the velocity which follows from our
definition. Whatever the value of $r$ (whence $v_\mathrm{prop}$),
$v_\mathrm{our}$ remains smaller than $c$.

The aim of this Paper is to derive a decomposition of the cosmological
redshift, so we have assumed that the distant galaxy lies on the past
light cone of the observer. However, both
Equations~(\ref{eq:U^0})--(\ref{eq:U^1}) of the parallel transport and
their initial conditions~(\ref{eq:init}) are completely independent of
this assumption. Consequently, in the solutions of these equations and
in the resulting Equation~(\ref{eq:v_rec-prop}), the proper coordinate
of a distant galaxy can be arbitrarily large; in particular, not
limited by the radius of the particle horizon. As a corollary, not
only the part of the cosmic substratum contained within the particle
horizon, but {\em entire\/} cosmic substratum expands
subluminally. Needless to say, the issue whether the Universe is open
or closed does not change anything: even in an open universe, at any
instant of time every particle has -- though allowably arbitrarily
large -- finite proper coordinate.

The only exception for subluminality of the expansion of the Universe
is the moment of initial singularity. As then $r \to 0$, it is better
to write $H r = \dot{a}x$. The early Universe is dominated by
ultra-relativistic particles and radiation so its scale factor evolves
as $a(t) \propto t^{1/2}$. Therefore, $\dot{a} \propto t^{-1/2}$, so
it diverges at $t = 0$. The only point in space which in spite of this
divergence is stationary at the Big-Bang is $x = 0$. This is obvious
because $x = 0$ is the origin of the coordinate system and all
velocities are measured relative to it. However, if the velocity of
the central observer is measured by another observer, non-zero
separation of the two observers and the fact that velocities are
relative imply that the velocity of the central observer relative to
{\em all\/} other observers is non-zero. (In order to prove it more
formally, use the FLRW metric centered on any other point.) Thus,
relative velocities of {\em all\/} FOs are non-zero and {\em at the
  initial singularity all velocities attain the velocity of light\/}.

Summing up, the expansion of the Universe is {\em never\/}
superluminal. A common misconception that the expansion is
superluminal is based on the wrong identification of recession
velocity with the `proper' recession velocity.

This subject is interesting in its own right and worth of further
study. Here we have presented it only briefly, to convince the reader
that a decomposition of the cosmological redshift into a gravitational
shift and a kinematic one for an arbitrary redshift is indeed
possible. This is so because for arbitrary redshifts, recession
velocities of galaxies are subluminal. We plan to study in more detail
the issue of (the lack of) superluminal expansion of the Universe on
superhorizon scales in a follow-up work.

\subsection{Recession velocities for specific models}
\label{sub:specific}
Figure~\ref{fig:transport} shows parallel transport of the velocity
four-vector of a distant galaxy, $\bfU$, from the galaxy to the
observer along the geodesic on the hypersurface of constant cosmic
time. The transport is plotted for two values of time: the observation
time, $\tO$ (dashed horizontal line), and the emission time, $\tE$
(solid horizontal line), both in units of $\tO$. The adopted
cosmological model is a matter-only, open universe with
$\Omega_\mathrm{m} = 0.24$. The observer is at the origin of the
coordinate system, $x = 0$, and the comoving coordinate of the
emitting galaxy is $x_\ast$ (vertical line) in units of $c
H_0^{-1}$. The adopted redshift of the source at $\tE$ (determining
the value of $x_\ast$) is equal to $3$. The transport is performed in
the comoving coordinates; in these coordinates all spatial components
of the galaxy's velocity four-vector are initially zero. During the
transport towards the observer, the radial component of the
four-velocity gradually increases. In the plot this component is
multiplied by $a(t)$, to enable a physical interpretation of the
transported vector, $\bfU(0)$, in the framework of special
relativity. Namely, at $\bfx = 0$ the scaled comoving components,
$(U^0, a U^1)$, are equal to the (non-scaled) components of $\bfU$ in
the {\em local inertial coordinates\/} of the observer,
$(U_\mathrm{in}^0, U_\mathrm{in}^1)$. Slanted long-dashed lines show
the direction of null vectors at the events $(\tO,0)$ and $(\tE,0)$ in
the latter coordinates. We see in Figure~\ref{fig:transport} that
$U_\mathrm{in}^1(0)$ [$= a U^1(0)$] is greater for $\tE$ than for
$\tO$. Thus, the recession velocity of the source at the time of
emission was greater than it is at present. This conclusion is
confirmed by the directions of the vectors: both vectors are timelike
(as required), but the direction of the vector at $(\tE,0)$ is closer
to the null direction than the direction of the corresponding vector
at $(\tO,0)$. It is expected, since a matter-only model constantly
decelerates, so its expansion must have been faster in the past.

In practice, it is more interesting to know the recession velocity of
a source {\em at the time it emitted the photons} which are reaching
the observer today. In other words, the velocities at the time of
emission are more interesting than the velocities at the time of
observation. (There is even no guarantee that the emitting galaxy
exists today.)

Combining Equation~(\ref{eq:v_rec}) with~(\ref{eq:redshift})
and~(\ref{eq:redshift-emission}), we obtain

\begin{equation}
\betaE = \tanh\left[\frac{ H(z)\, x(z)}{c\, (1+z)}\right] \,.
\label{eq:v_rec-em}
\end{equation}
Again, Equation~(\ref{eq:v_rec-em}) assures that the recession
velocity (at the time of emission) of a source with an arbitrarily
high redshift is subluminal ($\vE < c$). In the empty (Milne) model,
from~(\ref{eq:v_rec-em}) we obtain $\betaE = \tanh[\ln(1+z)]$, or

\begin{equation}
\betaE = \frac{(1+z)^2 -1}{(1+z)^2 +1} \,,
\label{eq:v_rec-empty}
\end{equation}
in agreement with the Doppler formula. In the EdS model we obtain

\begin{equation}
\betaE = \tanh\left\{2\!\left[(1+z)^{1/2} - 1\right]\right\} .
\label{eq:v_rec-EdS}
\end{equation}
In the currently favoured, flat model with non-zero $\Lambda$,
Equation~(\ref{eq:H(z)}) yields $H(z) = H_0\sqrt{1 - \Omega_\mathrm{m}
  + \Omega_\mathrm{m}(1+z)^3}$. The comoving distance is $x(z) = c
H_0^{-1} \! \int_0^z [1 - \Omega_\mathrm{m} +
  \Omega_\mathrm{m}(1+y)^3]^{-1/2} dy$.

Let us now consider a source located at a redshift, e.g., 6. For the
Milne model, the corresponding recession velocity at the time of
emission, $\vE^{\scriptscriptstyle(\mathrm{M})}$, is $0.96c$. For the
EdS model, this velocity, $\vE^{\scriptscriptstyle(\mathrm{E})}$, is
about $0.997c$. For the flat model with $\Lambda$, adopting
$\Omega_\mathrm{m} = 0.24$, the velocity,
$\vE^{\scriptscriptstyle(\mathrm{\Lambda})}$ equals approximately to
$0.991c$. We interpret these values in the following way. A Milne
universe is coasting, so $\vE^{\scriptscriptstyle(\mathrm{M})} =
\vO^{\scriptscriptstyle(\mathrm{M})}$. An EdS universe decelerates, so
it was expanding faster in the past. Therefore, it is logical that
$\vE^{\scriptscriptstyle(\mathrm{E})} >
\vE^{\scriptscriptstyle(\mathrm{M})}$.\footnote{An alternative
  explanation of this inequality follows from the results of
  sub-subsection~\ref{subsub:EdS-em}. In the EdS model there is a
  large gravitational blueshift, so for the total redshift to be equal
  to that in the Milne model, a Doppler shift must be much larger. In
  other words, $\vE^{\scriptscriptstyle(\mathrm{E})}$ must be much
  closer to the velocity of light than
  $\vE^{\scriptscriptstyle(\mathrm{M})}$.}  Similarly, one can also
expect that $\vE^{\scriptscriptstyle(\mathrm{E})} >
\vE^{\scriptscriptstyle(\mathrm{\Lambda})}$: in the flat $\Lambda$
model, the assumed value of $\Omega_\mathrm{m}$ is smaller than unity;
moreover, there is a late-time phase of acceleration due to the
cosmological constant. Interestingly, however, the value of
$\vE^{\scriptscriptstyle(\mathrm{\Lambda})}$ is greater than
$\vE^{\scriptscriptstyle(\mathrm{M})}$ and much closer to
$\vE^{\scriptscriptstyle(\mathrm{E})}$. Apparently, for such a high
redshift (and the adopted value $\Omega_\mathrm{m} =0.24$), the effect
of initial deceleration of the Universe on recession velocities
strongly prevails over the opposing effect of its later
acceleration. In fact this is not so surprising, since for
$\Omega_\mathrm{m} =0.24$ the flat model starts to accelerate only for
the redshift as small as about $0.5$.

On the other hand, for an arbitrarily high redshift there is always a
range of sufficiently small values of $\Omega_\mathrm{m}$, for which
$\vE^{\scriptscriptstyle(\mathrm{\Lambda})} <
\vE^{\scriptscriptstyle(\mathrm{M})}$. We have checked that for
$\Omega_\mathrm{m} = 0.24$, the critical redshift for which
$\vE^{\scriptscriptstyle(\mathrm{\Lambda})} =
\vE^{\scriptscriptstyle(\mathrm{M})}$ is about $1.6$.

\section{Discussion}
\label{sec:disc}
Chodorowski (2007a) noted that in conformally flat coordinates in
open, zero-$\Lambda$ models, {\em coordinate} velocities of all
galaxies are constant and subluminal. Lewis et al.\ (2007) pointed out
that even in these coordinates, the {\em proper}
velocities\footnote{In this case, the proper distance is measured on
  the hypersurface of constant conformal time.} of sufficiently
distant galaxies are superluminal. Lewis's et al.\ (2007) criticism
was fully justified, because coordinate velocities are
meaningless. However, in this Paper we have shown that the proper
velocity is {\em not} a real velocity of a galaxy, so the fact that
the former can be superluminal does not imply the same for the
latter. Indeed, basing on the definition of relative velocity given in
this Paper we have demonstrated that velocities of all particles of
the cosmic substratum are in fact {\em not} superluminal.

In a recent paper, Faraoni~(2010) studied the generation of the
cosmological redshift by an instantaneous expansion of otherwise
Minkowski spacetime. A source and an observer were assumed to be at
rest relative to each other both at the time of emission and
observation of photons. It was also assumed that the universe expands
only at a single instant of time $\tX$ and $\tE < \tX < \tO$.
Faraoni~(2010) showed that the induced redshift is given by the
canonical formula for all FLRW universes, $1+z = a(\tO)/a(\tE)$. On
the basis of this finding and the fact that the source and the
observer were initially and finally at rest, he deduced that in the
studied case the cosmological redshift is entirely gravitational. This
is in agreement with our results, as will be explained
below. Equations~(\ref{eq:v_rec})--(\ref{eq:redshift}) are valid for
{\em any} function $a(t)$, not necessarily determined by the FL
equations. The assumptions made by Faraoni imply that $\dot{a}(\tE) =
\dot{a}(\tO) = 0$ and from Equation~(\ref{eq:redshift}) one can see
that, both at the time of emission and observation, the Dopplerian
component of the cosmological redshift is zero. [Alternatively, from
  Equation~(\ref{eq:v_rec}) it follows that the recession velocity
  vanishes.] However, Faraoni's (2010) conclusion that the
cosmological redshift is {\em in general} gravitational is
incorrect. The quantity $\dot{a}$ vanishes only for closed FLRW models
and only for a single instant of time. More importantly, {\em our}
Universe is certainly expanding now, and whatever the exact values of
cosmological parameters, its expansion has been continuous in the
past. Therefore, the cosmological redshift must be partly Dopplerian.

BH9 transported the four-velocity of a distant galaxy along the {\em
  null} geodesic connecting the source and the observer. [The original
  idea dates back to Synge~(1960) and Narlikar~(1994)]. Instead, as
the path of parallel transport we have chosen the geodesic connecting
the source and the observer on the hypersurface of constant cosmic
time. Both definitions are equally justified mathematically: both are
intrinsic (i.e.\ they do not depend on any coordinate system) and
reduce to the SR definition of relative velocity if the two observers
are at the same point. The reason that as a path of transport we have
chosen a geodesic on the hypersurface of constant cosmic time was to
separate the effects of the curvature of the Universe from the effects
of its evolution. We wanted to define the recession velocity at a {\em
  well-specified instant of time}. Transporting along a null geodesic
provides $v_{\mathrm{BH}}$, which is a sort of time-average (over the
time interval between the times of emission and observation) of our,
{\em instantaneous\/} recession velocity. Indeed, it can be easily
verified that for small redshifts, hence also small time intervals,
$v_{\mathrm{BH}} = (\vE + \vO)/2$. As a consequence of their
definition of relative velocity, BH9 found that in any FLRW cosmology
and for any value of the redshift the cosmological redshift is related
to the transported velocity $v_{\mathrm{BH}}$ by the relativistic
Doppler formula.\footnote{Which means that, in particular, for small
  redshifts the formula of BH9 does not reduce to the non-relativistic
  limit of Bondi~(1947)} From this fact they deduced that the
cosmological redshift is entirely Dopplerian. Our interpretation of
the redshift is different: in a non-empty Universe there is
gravitational field, inducing a gravitational shift. Therefore, with
an exception of the empty model, the origin of the cosmological
redshift must be partly gravitational. Using our definition of
relative velocity, we have thoroughly proven this conjecture in the
present Paper. Thus, the difference between the finding of BH9 and
ours is yet another hint that $v_{\mathrm{BH}}$ is not the recession
velocity of a galaxy at a given instant of cosmic time. Using
Equations~(\ref{eq:small_velocity})
and~(\ref{eq:small_redshift__final}) it can be shown that for small
redshifts, $z = (\betaE + \betaO)/2 + [(\betaE +
  \betaO)/2]^2/2$. Comparing this equation with
Equation~(\ref{eq:small_redshift__final}) we see that the relation
between $z$ and the effective velocity $(\betaE + \betaO)/2$ is indeed
Dopplerian. But this effective velocity is just $v_{\mathrm{BH}}$.

Different definitions of relative velocity lead to different
interpretations of the cosmological redshift. However, the definition
of BH9 and ours imply the same very important conclusion: recession
velocities of all galaxies inside our particle horizon are
subluminal. This fact is perhaps more fundamental than the specific
value of recession velocity for a given redshift.

\section{Summary}
\label{sec:summ}
In this Paper we have decomposed the cosmological redshift into a
Doppler shift and a `rest', which we have interpreted as a
gravitational shift. In order to extract the kinematic component from
the cosmological redshift, we had to define properly the recession
velocity of a distant galaxy. This velocity is a relative velocity of
the galaxy and the observer, and in GR one cannot directly compare
vectors at widely separated points. First, one has to transport
parallely the four-velocity of the distant galaxy to the
observer. Next, the transported four-velocity is transformed to the
local inertial coordinates of the observer and then the recession
speed is typically extracted from the (transformed) radial component
of the four-velocity. In general, parallel transport depends on the
chosen path, but geodesics are special curves in GR and they seem to
be a very natural choice for the problem under consideration. As the
path, Synge (1960), Narlikar (1994) and BH9 adopted the {\em null}
geodesic of photons emitted towards the observer. For reasons
explained in Section~\ref{sec:disc}, we have chosen a different path:
the geodesic connecting the source with the observer on the
hypersurface of constant cosmic time. BH9 found that in any FLRW
cosmology and for any value of the redshift, the cosmological redshift
is related to the transported velocity, as defined by them,
$v_{\mathrm{BH}}$, by the relativistic Doppler formula. From this fact
they deduced that the cosmological redshift is entirely Dopplerian.
However, from the Principle of Equivalence (PoE) it follows that the
cosmological redshift must be partly gravitational. The explanation of
this apparent paradox is that $v_{\mathrm{BH}}$ is not a recession
velocity at a specific instant of time, but rather a sort of an
effective, time-averaged velocity.

In Section~\ref{sec:specific} we have presented analytical relations
between the cosmological redshift and its Doppler component for two
particularly simple cosmological models. In an empty universe, the
cosmological redshift is entirely kinematic. This is expected, because
in the empty model there is no gravitational field. In an Einstein-de
Sitter universe, at the time of emission of photons the Doppler shift
is greater that the total shift, so a gravitational shift is in fact a
blueshift. At the time of observation, the situation is different: for
large enough redshifts the gravitational component is dominant; in
particular, even the particle horizon is receding with a speed
smaller than that of light. This marked difference in the magnitude of
kinematic components of the total redshift at the time of emission and
at the time of observation is due to the fact that all matter-only
models have expanded faster in the past.

Using only the Principle of Equivalence, in Section~\ref{sec:PoE} we
have independently decomposed the (non-relativistic) cosmological
redshift into a Doppler shift and a `rest', and have shown that in
this regime the `rest' is induced by the {\em Newtonian\/}
gravitational potential. The obtained result is in agreement with a
second-order expansion of the exact formula derived here in the
framework of GR (Sec.~\ref{sec:specific}), as well as with the
classical formula of Bondi (1947) and equivalent findings of Peacock
(1999) and Gr{\o}n \& Elgar{\o}y (2007).

Last but not least, in Section~\ref{sec:rec} we have critically
revised a widespread opinion that the recession velocities of
(sufficiently) distant galaxies are superluminal. We have demonstrated
that the commonly used proper velocity is {\em not} a recession
velocity, but merely a radial component of the galaxy's velocity
four-vector (when expressed in proper coordinates). As such, it can
indeed attain superluminal values. On the contrary, the definition of
the recession velocity, presented in this Paper, naturally implies
that the velocity is subluminal for any value of the galaxy's
redshift. The definition of BH9 implies exactly the same conclusion. 
Basing on our definition we have also argued that the expansion of the
Universe is subluminal even on superhorizon scales. We plan to study
this issue more comprehensively in a future work.

\section*{Acknowledgements}
This work was partially supported by the Polish Ministry of Science
and Higher Education under grant N N203 0253 33, allocated for the
period 2007--2010.

\end{document}